\documentclass[letter,apj]{emulateapj-rtx4}
\usepackage{graphicx}
\usepackage{enumerate}
\usepackage{amssymb, amsmath}
\usepackage{mathtools}
\usepackage{natbib}
\usepackage{color}
\usepackage{hyperref}
\usepackage{ulem}
\usepackage{soul}
\def\jcap{JCAP}

\newcommand{\ias}{1}
\newcommand{\sagan}{2}

\begin{document}

\title{Constraints on MACHO Dark Matter from Compact Stellar Systems in Ultra-Faint Dwarf Galaxies}
\author{
Timothy D.~Brandt\altaffilmark{\ias,\sagan}
}

\altaffiltext{\ias}{Institute for Advanced Study, Einstein Dr., Princeton, NJ}
\altaffiltext{\sagan}{NASA Sagan Fellow}

\begin{abstract}
I show that a recently discovered star cluster near the center of the ultra-faint dwarf galaxy Eridanus~II provides strong constraints on massive compact halo objects (MACHOs) of $\gtrsim$5~$M_\odot$ as the main component of dark matter.  MACHO dark matter will dynamically heat the cluster, driving it to larger sizes and higher velocity dispersions until it dissolves into its host galaxy.  
The stars in compact ultra-faint dwarf galaxies themselves will be subject to the same dynamical heating; the survival of at least ten such galaxies places independent limits on MACHO dark matter of masses $\gtrsim$10~$M_\odot$.  Both Eri~II's cluster and the compact ultra-faint dwarfs are characterized by stellar masses of just a few thousand $M_\odot$ and half-light radii of 13~pc (for the cluster) and $\sim$30~pc (for the ultra-faint dwarfs).  These systems close the $\sim$20--100~$M_\odot$ window of allowed MACHO dark matter and combine with existing constraints from microlensing, wide binaries, and disk kinematics to rule out dark matter composed entirely of MACHOs from $\sim$10$^{-7}$~$M_\odot$ up to arbitrarily high masses.  
\end{abstract}

\maketitle

\section{Introduction} \label{sec:intro}

Dozens of ultra-faint dwarf galaxies have recently been discovered as satellites of the Galaxy and Andromeda, and as members of the Local Group (\citealt{McConnachie_2012} and references therein; \citealt{Koposov+Belokurov+Torrealba+etal_2015,Bechtol+Drlica-Wagner+Balbinot+etal_2015,Laevens+Martin+Bernard+etal_2015}). 
These satellites have luminosities as low as $\sim$1000~$L_\odot$, and total masses inside the half-light radius that are at least 1--3 orders of magnitude larger than their stellar masses (\citealt{Simon+Geha_2007}; \citealt{McConnachie_2012} and references therein).
Many of them host negligible amounts of gas; all are understood to be dominated by their dark matter content.  

Ultra-faint dwarf galaxies are excellent places to learn about dark matter.  They currently provide the best constraints on annihilating weakly interacting massive particles (WIMPs), ruling out the simplest thermal relic cross-sections for particle masses of tens of GeV \citep{Ackerman+Albert+Anderson+etal_2015}.  Dwarf galaxies have also been a source of tension with cosmological simulations: cold dark matter simulations have long overpredicted the abundance of massive satellite galaxies \citep{Klypin+Kravtsov+Valenzuela+etal_1999,Moore+Ghigna+Governato+etal_1999}.  This problem may be resolved through a combination of newly discovered dwarfs and the inclusion of baryonic physics in simulations \citep{Brooks+Zolotov_2014}, but has also been suggested as evidence for exotic forms of dark matter or modified gravity \citep{Lovell+Eke+Frenk+etal_2012}.

While the evidence for dark matter's existence is overwhelming \citep{Spergel+Verde+Peiris+etal_2003,Clowe+Bradavic+Gonzalez+etal_2006,Planck_2015}, the identity of the dark matter particles remains mysterious.  One intriguing possibility is that the dark matter consists of black holes formed in the early Universe.  Such massive compact halo objects \citep[MACHOs,][]{Griest_1991} could be detected in the halo of our Galaxy by gravitational microlensing \citep{Paczynski_1986}.  Microlensing surveys, however, have now ruled out MACHOs between $\sim$10$^{-7}$ and $\sim$30 $M_\odot$ as the dominant component of dark matter in our Galaxy \citep{Alcock+Allsman+Alves+etal_2001,Tisserand+LeGuillou+Afonso+etal_2007}.  At MACHO masses $\gtrsim$100~$M_\odot$, the existence of fragile, wide halo binaries constrains their abundance \citep{Chaname+Gould_2004,Yoo+Chaname+Gould_2004}, though these limits rely heavily on just a few systems \citep{Quinn+Wilkinson+Irwin+etal_2009}. \cite{Quinn+Wilkinson+Irwin+etal_2009} showed that one binary used by \cite{Yoo+Chaname+Gould_2004} to claim constraints for MACHOs $\gtrsim$20~$M_\odot$ is likely spurious, which removes the constraints for masses $\lesssim$200~$M_\odot$.  

At MACHO masses $\gtrsim$100~$M_\odot$, \cite{Hernandez+Matos+Sussman+etal_2004} showed that some dwarf galaxy cores would be dynamically relaxed, in tension with the relatively constant-density cores that have been inferred \citep{Sanchez-Salcedo+Reyes-Iturbide+Hernandez_2006,Goerdt+Moore+Read+etal_2006}.  MACHOs of very high mass ($\gtrsim$10$^7$~$M_\odot$) are also ruled out by the kinematics of the Galactic disk \citep{Lacey+Ostriker_1985}.  The only constraints on a population of MACHOs between $\sim$30~$M_\odot$ and $\sim$100~$M_\odot$, however, currently come from limits on spectral distortions of the cosmic microwave background (CMB): black holes may have accreted during the early Universe, leaving an imprint on the CMB \citep{Ricotti+Ostriker+Mack_2008}.  However, other authors have argued that these constraints may not be definitive \citep{Bird+Cholis+Munoz+etal_2016}.  \cite{Munoz+Kovetz+Dai+etal_2016} have shown that MACHOs of these masses may also be probed by lensed fast radio bursts (FRBs).

In this paper, I derive MACHO constraints from the compact stellar distributions of ultra-faint dwarf galaxies and, in particular, from the survival of a star cluster in Eridanus~II.  Eridanus~II
was discovered as part of the Dark Energy Survey \citep{Koposov+Belokurov+Torrealba+etal_2015,Bechtol+Drlica-Wagner+Balbinot+etal_2015}.  It has an absolute magnitude of $M_V = -7.1$ and a half-light radius of almost 300~pc, and lies just beyond the Galaxy's virial radius at a distance of $366 \pm 17$~kpc \citep{Crnojevic+Sand+Zaritsky_2016}.  Eri~II hosts a single star cluster of absolute magnitude $M_V = -3.5$ ($\sim$2000~$L_{\odot,V}$) and half-light radius $r_h = 13$~pc.  The star cluster appears to be nearly coincident with the galaxy's center.  Several other ultra-faint dwarfs have stellar masses of a few thousand $M_\odot$ and half-light radii of $\sim$30~pc \citep{Bechtol+Drlica-Wagner+Balbinot+etal_2015,Koposov+Belokurov+Torrealba+etal_2015,Koposov+Casey+Belokurov+etal_2015,Laevens+Martin+Bernard+etal_2015}; these galaxies provide independent MACHO constraints.

Eri~II is one of the few dwarf galaxies with a star cluster, but it is not unique in this respect.  The Fornax dwarf spheroidal galaxy has long been known to host globular clusters \citep{Baade+Hubble_1939,Hodge_1961}.  Its five known globular clusters range from 240~pc to 1.6~kpc in projected separation from the galaxy center, and from $\sim$$4 \times 10^4$$~M_\odot$ to $\sim$$3 \times 10^5$$~M_\odot$ in mass \citep{Mackey+Gilmore_2003}.  Dynamical friction should cause these clusters to spiral in towards the center of Fornax \citep{Tremaine_1976}; their current existence outside of its core places interesting constraints on the properties of Fornax's dark halo \citep{Sanchez-Salcedo+Reyes-Iturbide+Hernandez_2006,Goerdt+Moore+Read+etal_2006,Inoue_2011,Cole+Dehnen+Read+etal_2012}.  Other dwarf galaxies do host nuclear star clusters \citep{Georgiev+Hilker+Puzia+etal_2009}; these are invariably much more massive and more tightly bound than the cluster in Eri~II.

In this paper I show that the star cluster in Eri~II has important implications for MACHO dark matter, and that the population of compact ultra-faint dwarfs provides similar, independent limits.  The paper is organized as follows.  In Section \ref{sec:heating}, I apply the theory of collisional stellar systems to dynamical heating of the cluster by MACHOs.  Section \ref{sec:constraints} presents the constraints on MACHO dark matter from the cluster in Eri~II and from the population of compact ultra-faint dwarfs.  I discuss and conclude with Section \ref{sec:conclusions}.

\section{Heating of a Star Cluster by MACHOs} \label{sec:heating}

A star cluster is a dynamic environment where gravitational interactions lead to the exchange of energy between stars.  These interactions cause the system to dynamically relax; they may be approximated as diffusion terms using the Fokker-Planck equation \citep{Chandrasekhar_1943}.  This approximation has enabled modeling of cluster density and velocity distributions (\citealt{King_1966,Meylan_Heggie_1997} and references therein).  When a range of masses is present, stellar interactions lead to mass segregation, in which the most massive bodies have the most spatially compact distribution \citep{Spitzer_1969}.

Two-body interactions tend to equalize the mean kinetic energies of different mass groups at a given location.  In a system consisting of $>$1~$M_\odot$ MACHOs and stars, the stars will gain energy from the MACHOs; a compact stellar system will puff up.  This can be treated as a diffusion problem, with weak scatterings gradually changing each star's velocity.  The sum of the diffusion coefficients for the parallel and perpendicular components of the velocity describes the evolution of a star's kinetic energy.  Assuming an isotropic Maxwellian velocity distribution for the dark matter particles and a locally uniform dark matter density, the relevant diffusion coefficient is given by
\begin{equation}
D\left[ \left(\Delta v \right)^2 \right] = \frac{4\sqrt{2}\pi G^2 f_{\rm DM} \rho m_a \ln \Lambda}{\sigma} \left[ \frac{{\rm erf}(X)}{X} \right],
\label{eq:diffusion_BT}
\end{equation}
where $\ln \Lambda$ is the Coulomb logarithm, $m_a$ and $\sigma$ are the MACHO mass and velocity dispersion, $\rho$ is the total dark matter density, and $f_{\rm DM}$ is the fraction of dark matter in MACHOs of mass $m_a$ \citep{Binney+Tremaine_2008}.  The variable $X$ is the ratio of the stellar velocity to that of the MACHOs, $X \equiv v_*/(\sqrt{2}\sigma)$.  I will assume that the stars are relatively cold compared to the dark matter, $v_* \lesssim \sigma$, which implies that ${\rm erf}(X)/X \sim 1$.  This assumption is satisfied for all of the cluster and dark matter parameters used in the following section.  For the Coulomb logarithm, 
\begin{equation}
\ln \Lambda \approx \ln \left( \frac{b_{\rm max} \langle v^2 \rangle}{G(m + m_a)}\right) \approx \ln \left( \frac{r_h \sigma^2}{G(m + m_a)}\right),
\end{equation}
where $m$ and $m_a$ are the masses of the cluster stars and MACHOs, respectively, $\langle v^2 \rangle \approx \sigma^2$ is the typical relative velocity, and $b_{\rm max}$ is the maximum impact parameter \citep{Binney+Tremaine_2008}, which I take to be the half-light radius $r_h$.  For 10~$M_\odot$ MACHOs with $\sigma=10$~km\,s$^{-1}$ and $r_h=13$~pc, $\ln \Lambda \approx 10$.  
As usual, the Coulomb logarithm is very insensitive to the assumed halo properties.  

If dark matter is a mixture of MACHOs and low-mass particles like WIMPs, dynamical heating will compete with dynamical cooling.  The cooling rate from WIMPs is given by
\begin{equation}
D[\Delta E] = vD[\Delta v_{||}] = -\frac{4\pi G^2\rho v_* m_* (1 - f_{\rm DM}) \ln \Lambda}{\sigma^2}G(X),
\end{equation}
with $X \equiv v_*/(\sqrt{2}\sigma)$ as before, $m_*$ being the mass of a typical star, and 
\begin{equation}
G(X) = \frac{1}{2X^2} \left[ {\rm erf}(X) - \frac{2X}{\sqrt{\pi}} \exp{-X^2} \right].
\end{equation}
Dynamical heating will dominate over cooling by a factor
\begin{equation}
\frac{\rm heating}{\rm cooling} \sim \frac{m_a \sigma}{\sqrt{2} m_* v_*} \left( \frac{{\rm erf}(X)}{XG(X)} \right) \left( \frac{f_{\rm DM}}{1 - f_{\rm DM}} \right),
\label{eq:cooling}
\end{equation}
or a factor of $\sim$100$f_{\rm DM}/(1-f_{\rm DM})$ for 10~$M_\odot$ MACHOs.  Equation \eqref{eq:cooling} is {\it always} much larger than unity for the limiting $f_{\rm DM}$ derived in the following section; I neglect WIMP cooling for the rest of this paper.

Heating by MACHOs will add energy to the cluster, causing it to expand.  If the cluster is embedded in a dark matter core of constant density $\rho$, its potential energy per unit mass is given by
\begin{equation}
\frac{U}{M} = {\rm constant} -\alpha \frac{GM_*}{r_h} + \beta G \rho r_h^2,
\label{eq:energy}
\end{equation}
where $M_*$ is the cluster's stellar mass, $r_h$ is its projected half-mass radius, and $\alpha$ and $\beta$ are proportionality constants that depend on the mass distribution.  For a cored S\'ersic profile, $\alpha \sim 0.4$ and $\beta \sim 10$; I will adopt these values throughout the paper and assume them to remain constant as the cluster expands.  The measured profile of the cluster in the dwarf galaxy Eridanus~II is a S\'ersic profile with $n \approx 0.2$ \citep{Crnojevic+Sand+Zaritsky_2016}.  Performing the integrals using the python package \verb|sersic| \citep{Novak+Jonsson+Primack+etal_2012} gives values of $\alpha \approx 0.36$ and $\beta \approx 7.2$, which would make this cluster expand slightly faster than with my fiducial $\alpha$ and $\beta$.

\begin{figure}
\includegraphics[width=\linewidth]{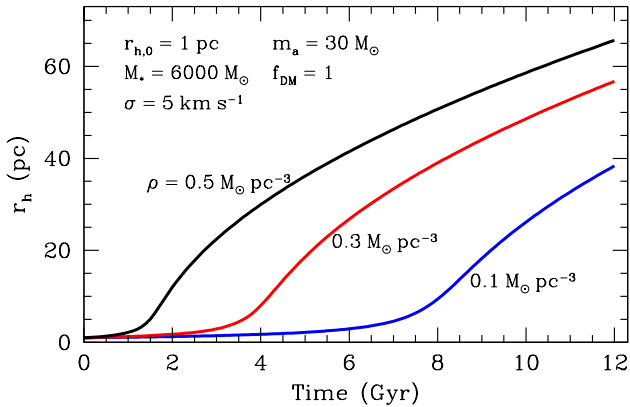}
\caption{Dynamical heating of a 6000~$M_\odot$ star cluster by 30~$M_\odot$ MACHOs at three fiducial densities, neglecting mass loss from the cluster.  The cluster expands slowly until its mean density equals that of the MACHOs, and then expands as $r_h \sim \sqrt{t}$.}
\label{fig:evol_rh}
\end{figure}

Using the virial theorem, $E_{\rm tot}=-\frac{1}{2}U$, Equations \eqref{eq:energy} and \eqref{eq:diffusion_BT} combine to give an implicit equation for the evolution of the half-light radius,
\begin{align}
\frac{dr_h}{dt} = \frac{4\sqrt{2}\pi G f_{\rm DM} m_a}{\sigma} \ln \Lambda \left(\alpha\frac{M_*}{\rho r_h^2} + 2\beta r_h \right)^{-1}.
\label{eq:evol_rh}
\end{align}
As long as the star cluster is dark-matter dominated, Equation \eqref{eq:evol_rh} is independent of the dark matter density.  A compact stellar system will expand slowly until it becomes dominated by its dark matter content, and then expand with $r_h \sim \sqrt{t}$.  Figure \ref{fig:evol_rh} demonstrates this behavior for a 6000~$M_\odot$ cluster with an initial half-light radius of 1~pc for 30~$M_\odot$ MACHOs at three fiducial dark matter densities, taking $\alpha=0.4$ and $\beta=10$.  

Motivated by Equation \eqref{eq:evol_rh}, I define two characteristic lifetimes for a stellar system.  The first is the time for it to puff up to its observed size from the $\sim$2~pc core radius of a typical Galactic star cluster \citep[][2010 edition]{Kharchenko+Piskunov+Roser+etal_2005,Harris_1996}.  The second is the timescale to double in area (increase by a factor of $\sqrt{2}$ in linear size).  In the limit of a dark-matter dominated system, these timescales are equal to each other, and to the timescale for the cluster to gain energy equal to its current kinetic energy.

\section{Constraints from the Ultra-Faint Dwarfs} \label{sec:constraints}

I now combine Equation \eqref{eq:evol_rh} with the observed survival of compact ultra-faint dwarf galaxies and of the star cluster in the core of Eridanus~II to constrain MACHO dark matter.  As described in the previous section, I define two characteristic lifetimes: (1) the time for the cluster to expand to its current size from the $\sim$2~pc core of a typical Galactic cluster; and (2) the time to double its current area.  By requiring these times, derived using Equation \eqref{eq:evol_rh}, to be longer than the cluster's age, I derive corresponding constraints on the abundance of MACHO dark matter.

\subsection{The Cluster in Eridanus II}

\begin{figure*}
\includegraphics[width=0.5\linewidth]{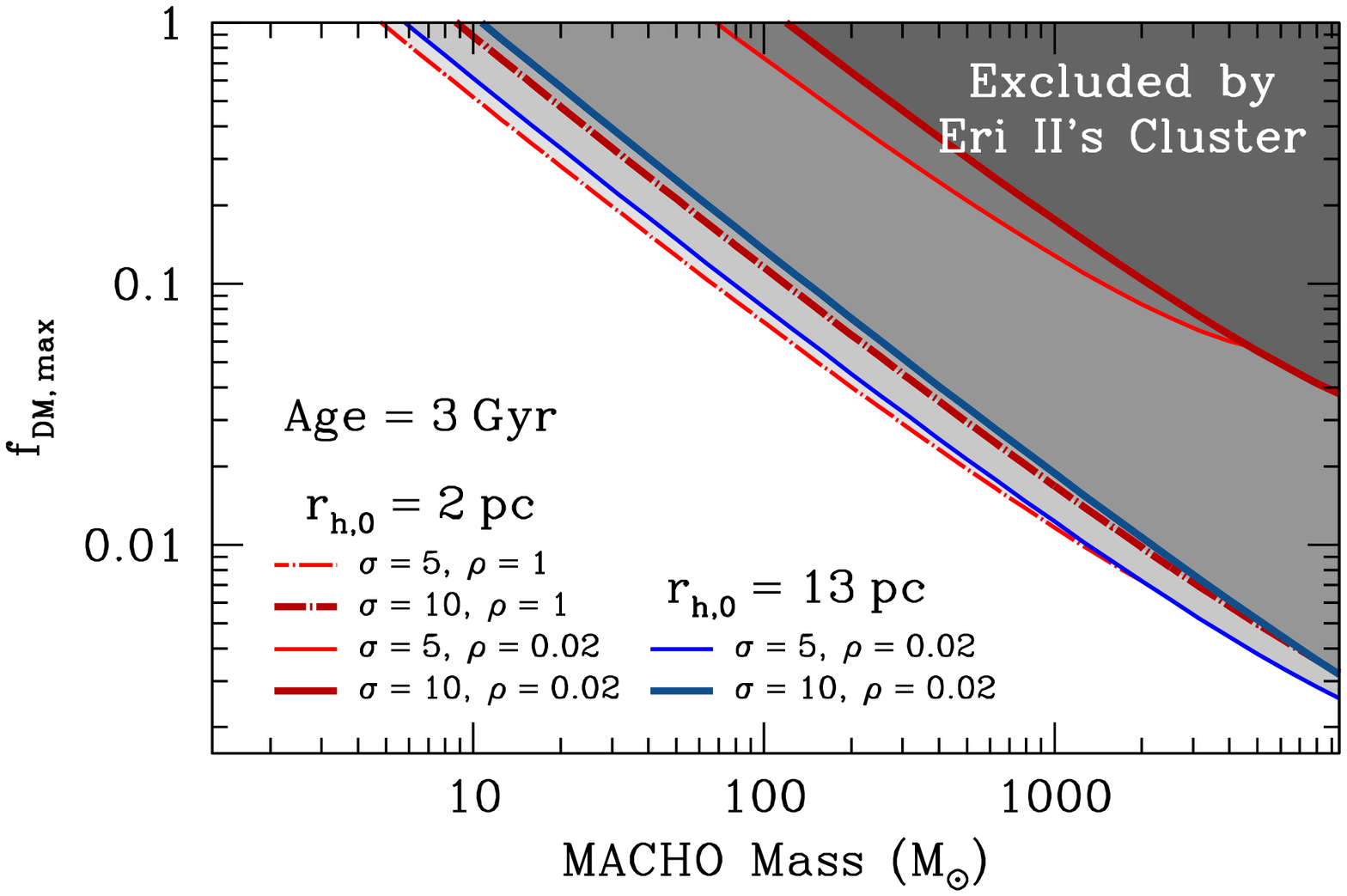}
\includegraphics[width=0.5\linewidth]{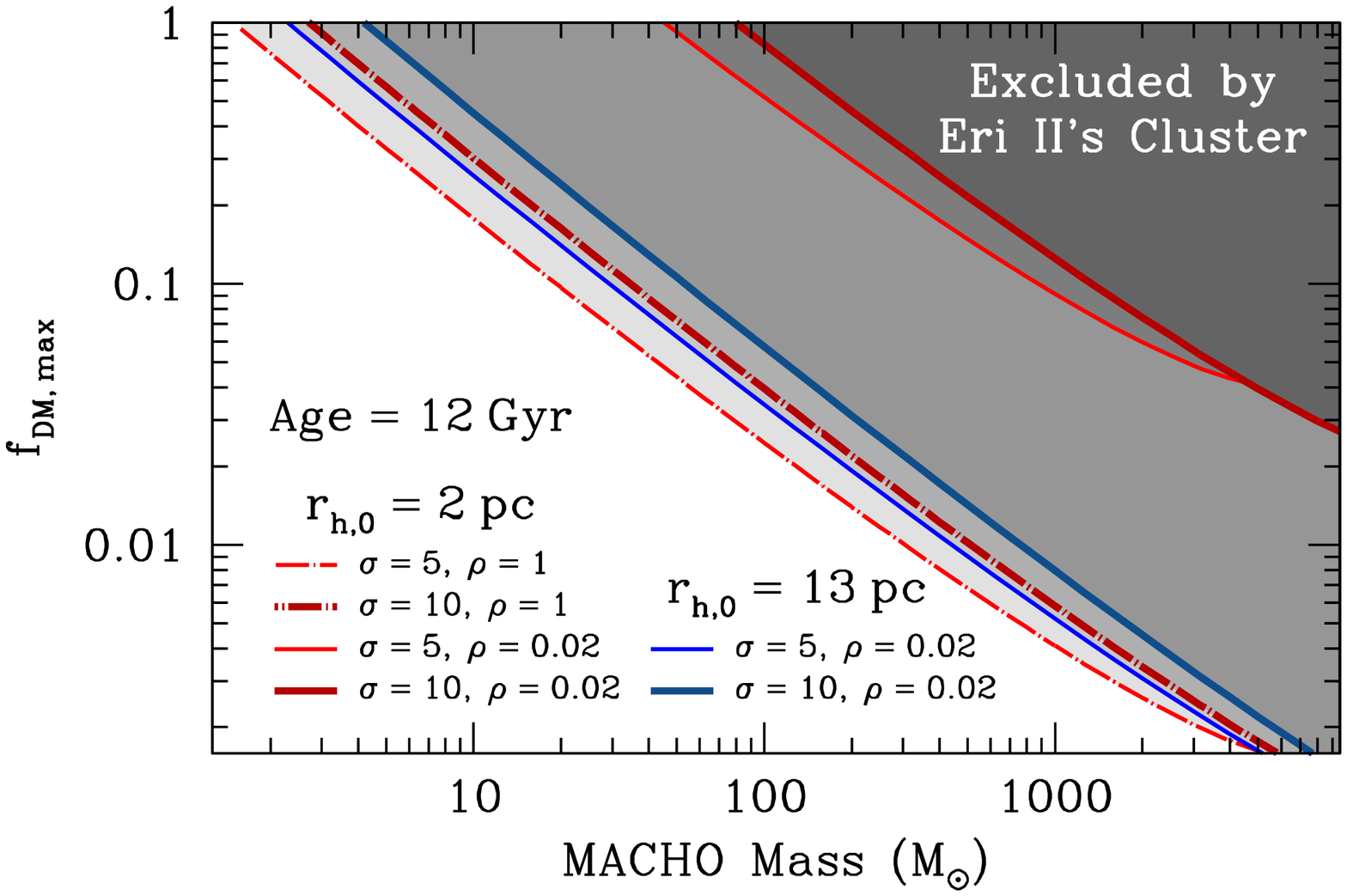}
\caption{MACHO constraints from the survival of the star cluster near the core of Eridanus II, assuming a cluster age of 3~Gyr (left panel) and 12~Gyr (right panel).  The units for the dark matter density $\rho$ and velocity dispersion $\sigma$, are $M_\odot\,$pc$^{-3}$ and km\,s$^{-1}$, respectively.  The limits come from requiring that the timescale to grow from $r_{h,0}= 2$~pc to the observed $r_h = 13$~pc is longer than the cluster age (red lines), or from requiring that the timescale to double in area (increase by $\sqrt{2}$ in $r_h$) is longer than the cluster age (blue lines).}
\label{fig:macho_constraints}
\end{figure*}

The star cluster in Eridanus~II is believed to be at least $\sim$3~Gyr old, and could be as old as $\sim$12~Gyr \citep{Crnojevic+Sand+Zaritsky_2016}.  At an age of 3~Gyr, the $V$-band mass-to-light ratio for a metal-poor stellar system is $\sim$1~$M_\odot/L_\odot$, while this ratio is $\sim$3~$M_\odot/L_\odot$ for an old system \citep{Maraston_2005}.  The cluster's observed $M_V=-3.5$ thus implies a stellar mass of $\sim$2000~$M_\odot$ at an age of 3~Gyr, or a mass of $\sim$6000~$M_\odot$ at an age of 12~Gyr.  The system has an observed half-light radius $r_h = 13$~pc.  I assume the system to have resided within the core of the dark matter halo for its entire life, and derive MACHO limits by requiring the timescales for dynamical heating to be longer than the cluster's age.

Figure \ref{fig:macho_constraints} shows the constraints for a range of plausible dark matter halo properties, with three-dimensional velocity dispersions of 5--10~km\,s$^{-1}$ and dark matter densities of 0.02--1~$M_\odot$\,pc$^{-3}$.  These values span the range of parameters characteristic of ultra-faint dwarf galaxies (\citealt{Simon+Geha_2007,McConnachie_2012} and references therein).  At an age of 3~Gyr (left panel), MACHOs $\lesssim$15~$M_\odot$ are excluded from making up all of the dark matter unless the Eri~II cluster was initially compact and remains embedded in a low-density, high-dispersion halo.  In this case, a cluster of the observed size is a transient phenomenon; similar objects should be rarer than compact low-mass clusters.  If the cluster has spent $\sim$12~Gyr near the center of its halo (right panel), the constraints strengthen.  

The preceding discussion assumed a roughly constant dark matter density profile (a core larger than the cluster).  Assuming a cuspy dark matter profile with the cluster at the dynamical center would strengthen the conclusions.  Such an assumption would make the cluster dominated by dark matter at a smaller half-light radius; it would quickly begin to evolve with $r_h \sim \sqrt{t}$ independently of dark matter density (Equation \eqref{eq:evol_rh}). Further, the velocity dispersion of the dark matter particles is expected to fall toward the center of an NFW halo \citep{Ferrer+Hunter_2013}.  Lower velocity dispersions would make MACHOs even more effective at dynamical heating, improving constraints on their abundance.  If, on the other hand, the cluster were slightly offset from the dynamical center of a strong dark matter cusp, it would be tidally shredded in a dynamical time.

\subsection{Constraints from Other Ultra-Faint Dwarfs} \label{subsec:ufdwarfs}

The entire stellar population of a dwarf galaxy will also be dynamically heated by MACHOs.  Many compact ultra-faint dwarf galaxies are now known, with stellar masses $\lesssim$3000~$M_\odot$ (assuming a mass-to-light ratio $M/L_V = 3$~$M_\odot/L_{\odot,V}$), half-light radii $\lesssim$30~pc, and central densities $\sim$1~$M_\odot$\,pc$^{-3}$.  Table \ref{tab:ufdwarfs} lists some basic properties of ten compact ultra-faint dwarfs (plus the star cluster in Eri~II); all but three were discovered since 2015.  Where measured, the ages of the stars are consistent with $\sim$10~Gyr \citep{Bechtol+Drlica-Wagner+Balbinot+etal_2015,Laevens+Martin+Bernard+etal_2015}.  The mean densities listed are vulnerable to different definitions of the half-light or half-mass radius, and should be treated as uncertain to at least a factor of $\sim$2.  The compact ultra-faint dwarfs constrain MACHO dark matter in the same way as the star cluster in Eri~II: I use the same two heating timescales and require one or the other to be longer than 10~Gyr.

\begin{deluxetable}{lccccr}
\tablecaption{Properties of Compact Ultra-Faint Dwarf Galaxies}
\tablehead{
    Name &
    $r_h$\tablenotemark{$\dagger$} &
    $L_V$ &
    $\overline{\rho}_{1/2}$ &
    $\sigma_*$ & 
    Ref.\tablenotemark{$\dagger\dagger$} \\
    & 
    pc &
    $L_\odot$ &
    $M_\odot$\,pc$^{-3}$ &
    km\,s$^{-1}$ &
    \phantom{*}
    }
\startdata
Wil I   & $25 \pm 6$     & 1000 & 4 & $4.3^{+2.3}_{-1.3}$ & 1,2 \\
Seg I   & $29^{+8}_{-5}$ & 300 & 3 & $3.9 \pm 0.8$ & 2,3 \\
Seg II  & $35 \pm 3$     & 900 & 1 & $3.4^{+2.5}_{-1.2}$ & 4 \\
Ret II  & $32^{+2}_{-1}$/$55^{+5}_{-5}$ & 1500\tablenotemark{*} & 2 & $3.2^{+1.6}_{-0.5}$ & 5,6,7 \\
Hor I   & $30^{+4}_{-3}$/$60^{+76}_{-30}$ & 2000\tablenotemark{*} & 5 & $4.9^{+2.8}_{-0.9}$ & 5,6,7 \\
Pic I   & $29^{+9}_{-4}$/$43^{+153}_{-21}$ & 2000\tablenotemark{*} & & & 5,6 \\
Pho II  & $26^{+6}_{-4}$/$33^{+20}_{-11}$ & 1500\tablenotemark{*}  & & & 5,6   \\
Ind I   & $37^{+13}_{-8}$/$12^{+2}_{-2}$ & 1000\tablenotemark{*} &   & & 5,6 \\
Eri III & $14^{+13}_{-3}$/$11^{+8}_{-5}$ & 500\tablenotemark{*} &   & & 5,6 \\
Dra II  & $19^{+8}_{-6}$ & 1000 &  & & 8 \\
Eri II\tablenotemark{**} & $13 \pm 1$ & 2000 & & & 9
\enddata
\tablenotetext{$\dagger$}{Where two values are given, the first is from \cite{Koposov+Belokurov+Torrealba+etal_2015} and the second from \cite{Bechtol+Drlica-Wagner+Balbinot+etal_2015}.}
\tablenotetext{$\dagger\dagger$}{References abbreviated as: 
1 \citep{Martin+Ibata+Chapman+etal_2007}; 
2 \citep{Martin+deJong+Rix_2008}; 
3 \citep{Simon+Geha+Minor+etal_2011};
4 \citep{Belokurov+Walker+Evans+etal_2009}; 
5 \citep{Bechtol+Drlica-Wagner+Balbinot+etal_2015}; 
6 \citep{Koposov+Belokurov+Torrealba+etal_2015}; 
7 \citep{Koposov+Casey+Belokurov+etal_2015}; 
8 \citep{Laevens+Martin+Bernard+etal_2015};
9 \citep{Crnojevic+Sand+Zaritsky_2016} 
}
\tablenotetext{*}{Geometric means of \cite{Koposov+Belokurov+Torrealba+etal_2015} and \cite{Bechtol+Drlica-Wagner+Balbinot+etal_2015}, rounded to $500~L_\odot$.}
\tablenotetext{**}{Values are for the central star cluster only.}
\label{tab:ufdwarfs}
\end{deluxetable}

\begin{figure}
\includegraphics[width=\linewidth]{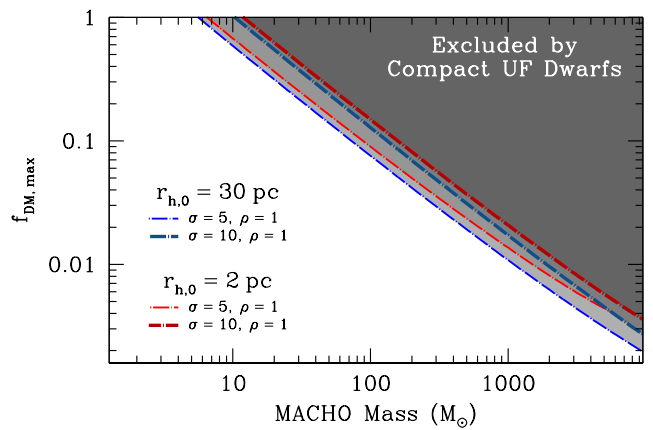}
\caption{MACHO constraints from the observed sizes of compact ultra-faint dwarf galaxies, assuming a stellar mass of $3000$~$M_\odot$, a current half-light radius $r_h=30$~pc, and an age of 10~Gyr. The units for the dark matter density $\rho$ and velocity dispersion $\sigma$, are $M_\odot\,$pc$^{-3}$ and km\,s$^{-1}$, respectively.  The limits come from requiring that the timescale to grow from $r_{h,0}= 2$~pc to $r_h = 30$~pc is longer than 10~Gyr (red lines), or from requiring that the timescale to double in area (increase by $\sqrt{2}$ in $r_h$) is longer than 10~Gyr (blue lines).}
\label{fig:ufdwarfs}
\end{figure}

Figure \ref{fig:ufdwarfs} shows the limits on MACHO dark matter implied by a fiducial compact ultra-faint dwarf, with $r_h=30$~pc, $M_*=3000$~$M_\odot$, and a central dark matter density $\rho=1$~$M_\odot$\,pc$^{-3}$, for three-dimensional velocity dispersions of 5 and 10~km\,s$^{-1}$.  The observed ultra-faint dwarfs lie within this range; with one-dimensional velocity dispersions between 3 and 6~km\,s$^{-1}$ (Table \ref{tab:ufdwarfs}).  The survival of the compact ultra-faint dwarfs listed in Table \ref{tab:ufdwarfs} rules out dark matter consisting entirely of MACHOs of mass $\gtrsim$10~$M_\odot$.

\section{Discussion and Conclusions} \label{sec:conclusions}

\begin{figure}
\includegraphics[width=\linewidth]{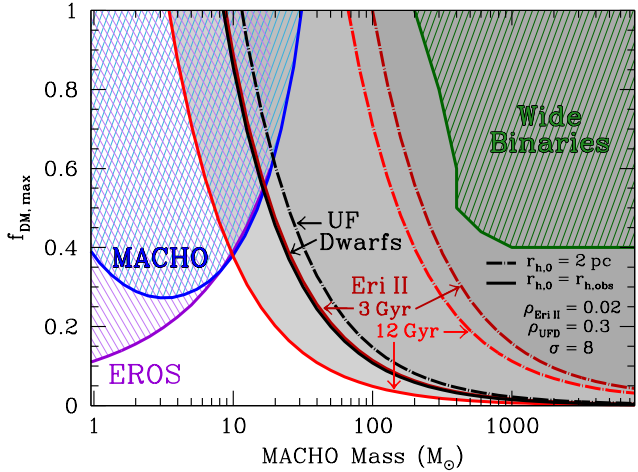}
\caption{Constraints on MACHO dark matter from microlensing \citep[blue and purple,][]{Alcock+Allsman+Alves+etal_2001,Tisserand+LeGuillou+Afonso+etal_2007} and wide Galactic binaries \citep[green,][]{Quinn+Wilkinson+Irwin+etal_2009}, shown together with the constraints from the survival of compact ultra-faint dwarf galaxies and the star cluster in Eridanus~II.  I conservatively adopt a dark matter density of $0.02$~$M_\odot$\,pc$^{-3}$ in Eri~II and $0.3$~$M_\odot$\,pc$^{-3}$ in the ultra-faint dwarfs, assume a three-dimensional velocity dispersion $\sigma=8$~km\,s$^{-1}$, and use two definitions of the heating timescale.  A low-density halo and initially compact cluster weaken the constraints from Eri~II.  Even in this case, assuming dark matter halos to have the properties that are currently inferred, MACHO dark matter is excluded for all MACHO masses $\gtrsim$10$^{-7}$~$M_\odot$.}
\label{fig:all_constraints}
\end{figure}

The star cluster in the core of the newly discovered dwarf galaxy Eridanus~II provides strong constraints on a region of MACHO parameter space difficult to probe with either microlensing or wide Galactic binaries; the population of compact, ultra-faint dwarfs provides similar, independent limits.  Figure \ref{fig:all_constraints} compares the constraints derived in Section \ref{sec:constraints} using conservative assumptions about the dark matter halos to constraints from microlensing \citep{Alcock+Allsman+Alves+etal_2001,Tisserand+LeGuillou+Afonso+etal_2007} and wide Galactic halo binaries \citep{Quinn+Wilkinson+Irwin+etal_2009}.  The kinematics of the Galactic disk provide an independent limit on the abundance of very massive ($\gtrsim$10$^7$~$M_\odot$) MACHOs \citep{Lacey+Ostriker_1985}.  For dark matter halos consistent with measured dwarf properties (Table \ref{tab:ufdwarfs}), MACHO dark matter is ruled out over the entire open region of masses.  If Eri~II's cluster is old and was initially puffier than Galactic clusters, it provides especially strong limits.

While Eri~II's cluster likely provides the best limits on MACHOs from $\sim$10~$M_\odot$ up to thousands of $M_\odot$, there are ways to evade its constraints.  The cluster, for example, could have recently spiraled into the center of Eri~II due to dynamical friction, having spent most of its life as a compact cluster in a low-density MACHO environment.  However, the inspiral timescale is inversely proportional to the cluster mass \citep{Binney+Tremaine_2008}, and the cluster in Eri II is 1.5--2 orders of magnitude less massive than Fornax 4 \citep{Mackey+Gilmore_2003}, the Fornax globular cluster nearest the center of that dwarf (at 240~pc in projected separation).  This scenario therefore requires very different dark matter halos in the two galaxies or severe mass loss during Eri~II's inspiral, and also luck to catch the cluster on the point of disruption.  This problem of coincidence is generic to any scenario in which Eri~II's cluster was initially compact.  The probability of observing the system in such a transient state is significantly higher if the cluster's age is $\sim$3~Gyr rather than $\sim$12~Gyr.

Other possibilities to evade the constraints include an intermediate-mass black hole ($\gtrsim$10$^4$~$M_\odot$) to provide binding energy, or a chance alignment such that the cluster only appears to reside in the center of Eri~II.  Both would be surprising.  Such a black hole would have a mass comparable to the total stellar mass of its host galaxy.  A massive black hole would also be expected to host a relaxed MACHO cluster of comparable mass, in which case it may not avoid the problem of dynamical heating at all.  A chance alignment of a cluster physically located at the galaxy's half-light radius is possible; the most na\"ive estimate, the fraction of solid angle lying within a few $r_h$ in projection, gives a chance alignment probability of $\sim$1\% at a physical distance of $\sim$300~pc from the galaxy core.  

While many scenarios could, in principle, account for the survival of the star cluster in Eri~II, it is harder to appeal to coincidence for the entire sample of compact ultra-faint dwarfs.  Assuming the measured velocity dispersions to reflect the properties of their dark matter halos, these dwarfs should have much larger half-light radii if their dark matter is all in the form of MACHOs $\gtrsim$10~$M_\odot$.   The strongest constraints, however, may come from the cluster in Eri~II, and could be improved with better data.  Precise photometry with the {\it Hubble Space Telescope} could resolve the question of whether the cluster is intermediate-age or old, while spectroscopy of cluster members and nonmembers would give another probe of Eri~II's dark matter content.  While future observations will determine the strength of the constraints from Eri~II, existing data from Eri~II and from the sample of compact ultra-faint dwarfs appear sufficient to rule out dark matter composed exclusively of MACHOs for all masses above $\sim$10$^{-7}$~$M_\odot$.

\acknowledgments{I thank Ben Bar-Or, Juna Kollmeier, Kris Sigurdson, and especially Scott Tremaine for helpful conversations and suggestions, and an anonymous referee for helpful comments.  This work was performed under contract with the Jet Propulsion Laboratory (JPL) funded by NASA through the Sagan Fellowship Program executed by the NASA Exoplanet Science Institute.}

\end{document}